\documentclass[prd,twocolumn,preprintnumbers,superscriptaddress,nofootinbib]{revtex4}
\usepackage[a4paper, hdivide={1.91cm,,1.165cm}, vdivide={1.83cm,,3.8cm}]{geometry}

\usepackage{amsmath}
\usepackage{graphicx}
\usepackage{xspace}
\usepackage{color}
\usepackage{units}
\usepackage{slashed}
\usepackage[hyperfootnotes=false]{hyperref}

\def \L {\mathcal{L}} 

\def \epsilon {\varepsilon} 

\def \vec#1{{\boldsymbol{#1}}}
\newcommand{\matrixx}[1]{\begin{pmatrix} #1 \end{pmatrix}} 

\newcommand{\Br}{\text{Br}}

\allowdisplaybreaks

\begin{document}

\title{Lepton flavor violation with light vector bosons}

\preprint{ULB-TH/16-02}

\author{Julian \surname{Heeck}}
\email{Julian.Heeck@ulb.ac.be}
\affiliation{Service de Physique Th\'eorique, Universit\'e Libre de Bruxelles, Boulevard du Triomphe, CP225, 1050 Brussels, Belgium}

\hypersetup{
    pdftitle={Lepton flavor violation with light vector bosons},
    pdfauthor={Julian Heeck}
}


\begin{abstract}
New sub-GeV vector bosons with couplings to muons but not electrons have been discussed in order to explain the muon's magnetic moment, the gap of high-energy neutrinos in IceCube or the proton radius puzzle. If such a light $Z'$ not only violates lepton universality but also lepton flavor, as expected for example from the recent hint for $h\to\mu\tau$ at CMS, the two-body decay mode $\tau \to \mu Z'$ opens up and for $M_{Z'} < 2 m_\mu$ gives better constraints than $\tau\to 3\mu$ already with 20-year-old ARGUS limits. We discuss the general prospects and motivation of light vector bosons with lepton-flavor-violating couplings.
\end{abstract}

\maketitle


\section{Introduction}

The search for physics beyond the Standard Model (SM) follows two complementary paths: on the one hand we have the high-energy-frontier experiments, most importantly ATLAS and CMS at the LHC, which can probe new particles with TeV-scale masses if they interact sufficiently strong; on the other hand we have the high-precision frontier, which aims at minute deviations in low-energy observables.
New physics can typically fall into either regime depending on the parameters involved; for example, a new gauge boson $Z'$ which acquires its mass $M_{Z'}$ from a TeV-scale vacuum expectation value $\langle \phi\rangle = M_{Z'}/g'$ can be discovered at the LHC for gauge couplings $g'= \mathcal{O}(1)$. For smaller gauge couplings, say $g' = 10^{-6}$, the same model would however only be testable at the precision frontier, which can probe MeV-scale masses.
If the $Z'$ has flavor-violating couplings, the strongest constraints typically arise from rare decays such as $\mu\to e\gamma$. Studies are usually concerned with very heavy $Z'$ masses~\cite{Langacker:2000ju,Murakami:2001cs,Chiang:2011cv}, even though one can again consider rather light $Z'$ with a small coupling constant. It is precisely this region of parameter space we are interested in here. (Similar studies can be (and have been) performed for light (pseudo-)scalar bosons, e.g.~Majorons, axions or familions~\cite{Wilczek:1982rv,Grinstein:1985rt,Feng:1997tn,Hirsch:2009ee,Jaeckel:2013uva,Celis:2014iua,Celis:2014jua}.)

Motivations for a light $Z'$ are plentiful. There is the long-standing $\sim 3\sigma$ discrepancy $a_\mu^\mathrm{exp}-a_\mu^\mathrm{SM} = (236\pm 87)\times 10^{-11}$~\cite{Crivellin:2015hha} of the muon's anomalous magnetic moment $a_\mu \equiv (g-2)_\mu/2$, which can be resolved with a sufficiently muon-philic $Z'$~\cite{Gninenko:2001hx,Baek:2001kca,Carone:2013uh} with mass below GeV~\cite{Altmannshofer:2014pba}.\footnote{Note that the lepton-universal ``hidden photon'' solution of $(g-2)_\mu$ has recently been excluded experimentally~\cite{Lees:2014xha,Batley:2015lha}.}
The same light $Z'$ can explain the gap of high-energy neutrinos in IceCube~\cite{Araki:2014ona,DiFranzo:2015qea,Araki:2015mya, Kamada:2015era}.
An MeV-scale $Z'$ has also been proposed to resolve the proton radius puzzle~\cite{Pohl:2010zza,Pohl:2013yb}, which requires couplings to protons and muons~\cite{TuckerSmith:2010ra}.

Interestingly, the $Z'$ violates lepton \emph{universality} in all cases in order to avoid strong bounds on electron couplings. (A popular UV-complete model is based on gauged $U(1)_{L_\mu-L_\tau}$, which is not only free of anomalies~\cite{He:1990pn,Foot:1990mn,He:1991qd} but also motivated by neutrino mixing angles~\cite{Binetruy:1996cs,Bell:2000vh,Choubey:2004hn}.)
From lepton non-universality it is but a small stretch to imagine lepton-flavor violating (LFV) couplings of the light $Z'$.
This holds true in particular considering the tantalizing $2.5\sigma$ hint for the LFV scalar decay $h\to\mu\tau$ at CMS~\cite{Khachatryan:2015kon} and ATLAS~\cite{Aad:2015gha}, which has been shown in Ref.~\cite{Heeck:2014qea} to fit perfectly into a $U(1)_{L_\mu-L_\tau}$ model with LFV $Z'$ couplings.
Here we show that by simply taking the gauge coupling $g'$ to be small, the very same model can resolve the muon's magnetic moment and lead to a large LFV decay rate $\tau\to \mu Z'$.

We will focus on $Z'$ couplings to muons and taus, heavily inspired by $U(1)_{L_\mu-L_\tau}$ models and the observation that all experimental hints for lepton non-universality or LFV reside in the muon sector. An additional coupling to quarks can lead to further interesting effects and can be readily constructed, see e.g.~Refs.~\cite{Altmannshofer:2014cfa,Crivellin:2015mga,Crivellin:2015lwa}. While a light $Z'$ can not resolve the accumulating anomalies in $B$-meson decays~\cite{Fuyuto:2015gmk}, it can lead to non-standard neutrino interactions, as recently discussed in Ref.~\cite{Farzan:2015doa,Farzan:2015hkd}.
One sure source of additional couplings is kinetic mixing of our $Z'$ with the photon, which unavoidably arises in models with several $U(1)$ gauge groups~\cite{Galison:1983pa}. We will neglect this coupling here and refer to Ref.~\cite{Essig:2013lka} for a thorough discussion.

\section{Couplings}

We consider flavor-violating $Z'$ couplings in the $\mu$--$\tau$ sector as parametrized by the effective Lagrangian
\begin{align}
\begin{split}
\L =  \left(\overline{\mu},\,\overline{\tau}\right) \slashed{Z'} \left[\matrixx{v_{\mu\mu} & v_{\mu\tau} \\ v_{\mu\tau} & v_{\tau\tau}} - \matrixx{a_{\mu\mu} & a_{\mu\tau} \\ a_{\mu\tau} & a_{\tau\tau}}\gamma_5\right] \matrixx{\mu\\\tau}  .
\end{split}
\label{eq:lagrangian}
\end{align}
We will assume these couplings to be real and often collectively denote $v_{\alpha\beta}$ and $a_{\alpha\beta}$ as $g_{\alpha\beta}$.
Typically one would assume the off-diagonal entries to be generated from a small rotation, so that $|g_{\alpha\beta}|\ll |g_{\alpha\alpha}|$~\cite{Heeck:2011wj, Heeck:2014qea}. This does not necessarily have to be the case, as one can also build models with purely off-diagonal entries~\cite{Foot:1994vd}. We remain agnostic about the origin for now, and furthermore do not introduce a coupling to electrons in order to simplify the discussion. We do, however, expect a coupling to the neutrinos, which will be relevant for the neutrino trident production (NTP) bound of the $(g-2)_\mu$ resolution~\cite{Altmannshofer:2014pba}. Without introducing right-handed neutrinos, $SU(2)_L$ gauge invariance enforces the neutrino couplings
\begin{align}
\begin{split}
\L = \left(\overline{\nu}_\mu,\,\overline{\nu}_\tau\right) \slashed{Z'} \matrixx{v_{\mu\mu}+a_{\mu\mu} & v_{\mu\tau} +a_{\mu\tau} \\ v_{\mu\tau}+ a_{\mu\tau} & v_{\tau\tau}+a_{\tau\tau}} P_L \matrixx{\nu_\mu\\\nu_\tau} ,
\end{split}
\label{eq:nulagrangian}
\end{align}
with left-handed projector $P_L = (1-\gamma_5)/2$.
Right-handed neutrinos introduce a model dependence, but could have the same $U(1)'$ charges as the other leptons, e.g.~in $L_\mu-L_\tau$ models. In that case one has to assume neutrinos to be Majorana particles, because Dirac neutrinos coupled to a light $Z'$ would severely modify Big Bang nucleosynthesis (BBN) and exclude our region of interest to explain $(g-2)_\mu$~\cite{Heeck:2014zfa}.
Majorana masses can be obtained easily enough by means of a seesaw mechanism at the scale $M_{Z'}/g'$, see e.g.~Refs.~\cite{Heeck:2011wj, Heeck:2014qea}, which makes the heavy seesaw partners of the neutrinos unimportant for our study.

\subsection{Flavor conserving processes}

\begin{figure}[t]
\includegraphics[width=0.48\textwidth]{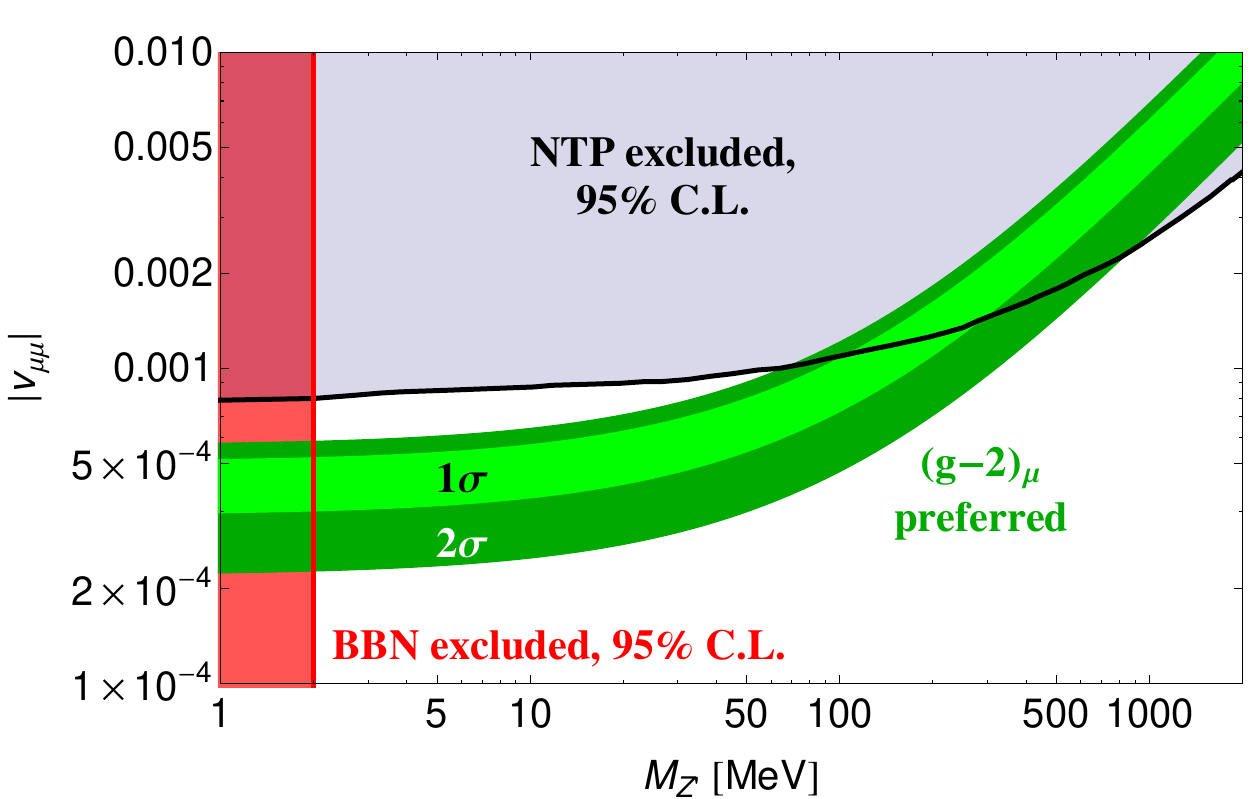}
\caption{
Limits on a gauge boson $Z'$ with only vector-like couplings $v_{\mu\mu}$ to muons and muon-neutrinos. The (light) green area is the preferred ($1\sigma$) $2\sigma$ region to resolve the muon's anomalous magnetic moment, 
gray is the NTP bound ($\nu_\mu N \to \nu_\mu N \mu^+\mu^-$) from CCFR~\cite{Altmannshofer:2014pba}, 
and red the BBN bound corresponding to $\Delta N_\mathrm{eff}\leq 1$ from $\nu\bar\nu\to Z'$ production~\cite{Ahlgren:2013wba}.
}
\label{fig:muonforce}
\end{figure}

Let us first review the effects of $g_{\alpha\beta}$ on flavor conserving processes.
The most important constraint---and motivation---for this kind of model comes from the muon's magnetic moment. The one-loop contribution can be readily calculated following e.g.~Ref.~\cite{Leveille:1977rc}.
In the limit $M_{Z'}\gg m_\mu$, we find the simple expression
\begin{align}
\Delta a_\mu &\simeq \frac{ v_{\mu\mu}^2 + v_{\mu\tau}^2 \left( \frac{3 m_\tau}{m_\mu}-2\right)- 5 a_{\mu\mu}^2 -a_{\mu\tau}^2 \left( \frac{3m_\tau}{m_\mu}+2\right) }{12\pi^2 M_{Z'}^2/m_\mu^2} ,
\end{align}
whereas the opposite limit $M_{Z'}\ll m_\mu$ gives
\begin{align}
\begin{split}
\Delta a_\mu \simeq &+\frac{v_{\mu\mu}^2}{8\pi^2}+ \frac{v_{\mu\tau}^2}{16\pi^2}\frac{ m_\mu m_\tau }{ M_{Z'}^2} \left(1-\frac{5m_\mu}{3m_\tau}\right)\\
&-\frac{a_{\mu\mu}^2}{4\pi^2}\frac{ m_\mu^2 }{ M_{Z'}^2}- \frac{a_{\mu\tau}^2}{16\pi^2}\frac{ m_\mu m_\tau }{ M_{Z'}^2} \left(1+\frac{5m_\mu}{3m_\tau}\right) .
\end{split}
\end{align}
A positive contribution---in order to resolve the discrepancy---requires \emph{vector} couplings, so we will mostly neglect $a_{\alpha\beta}$ in the following. (Constraints on $a_{\mu\mu}$ from parity violation have been derived in Ref.~\cite{Karshenboim:2014tka}.) See Fig.~\ref{fig:muonforce} for the preferred region in the $v_{\mu\mu}$--$M_{Z'}$ plane.
Note that all contributions except for the diagonal vector coupling are enhanced by $1/M_{Z'}^2$ in the limit of light $Z'$.

As pointed out in Refs.~\cite{Altmannshofer:2014cfa,Altmannshofer:2014pba}, any muon-philic $Z'$ solution to $(g-2)_\mu$ is constrained by measurements of NTP $\nu_\mu N \to \nu_\mu N \mu^+\mu^-$, most importantly from CCFR~\cite{Mishra:1991bv}. This essentially excludes the $Z'$ mass range above $\unit[900]{MeV}$ for a solution of $(g-2)_\mu$, see Fig.~\ref{fig:muonforce}. (We use an updated and fairly conservative value for $(g-2)_\mu$ from Ref.~\cite{Crivellin:2015hha}, so our numerical values differ from e.g.~Ref.~\cite{Altmannshofer:2014pba}.)
The remaining parameter space of interest for $(g-2)_\mu$ can be covered with future searches~\cite{Gninenko:2014pea}.

A very light $Z'$ will contribute to the relativistic degrees of freedom in the early Universe and severely modify BBN. Even if the $Z'$ couples only to neutrinos, one can obtain $95\%$~C.L.~limits $M_{Z'} > \unit[2]{MeV}$ for $g_{\alpha\alpha} > 10^{-9}$~\cite{Ahlgren:2013wba}. (We will not entertain the possibility of even smaller couplings and masses~\cite{Heeck:2010pg}.) Similar qualitative conclusions have been obtained in Refs.~\cite{Kamada:2015era,Farzan:2015hkd}. 
Kaon decay $K^-\to\mu^-\bar\nu Z'$ followed by the invisible decay $Z'\to\nu\nu$ gives additional constraints in particular on the axial couplings $a_{\alpha\beta}$~\cite{Barger:2011mt,Laha:2013xua}, but are not relevant here.

\subsection{Flavor violating processes}

\begin{figure*}[t]
\includegraphics[width=0.43\textwidth]{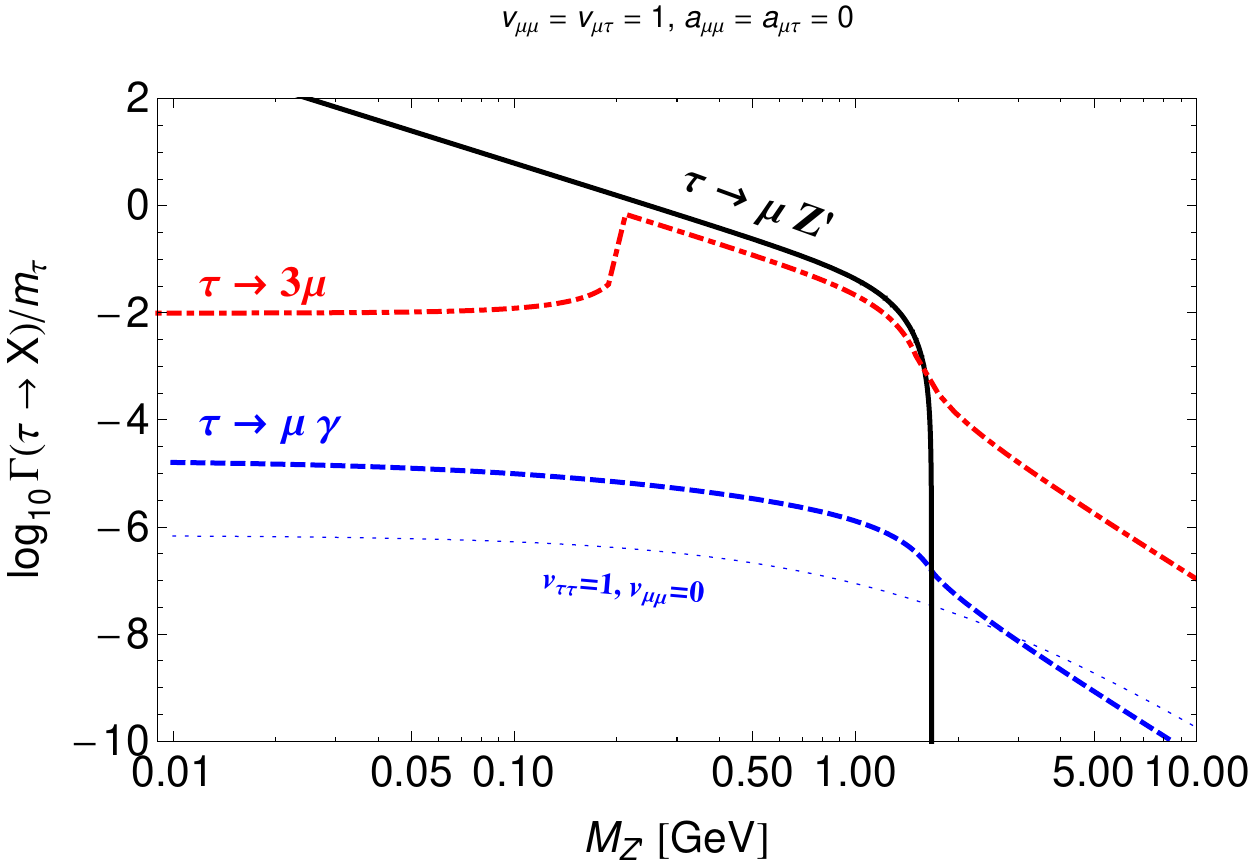}
\includegraphics[width=0.43\textwidth]{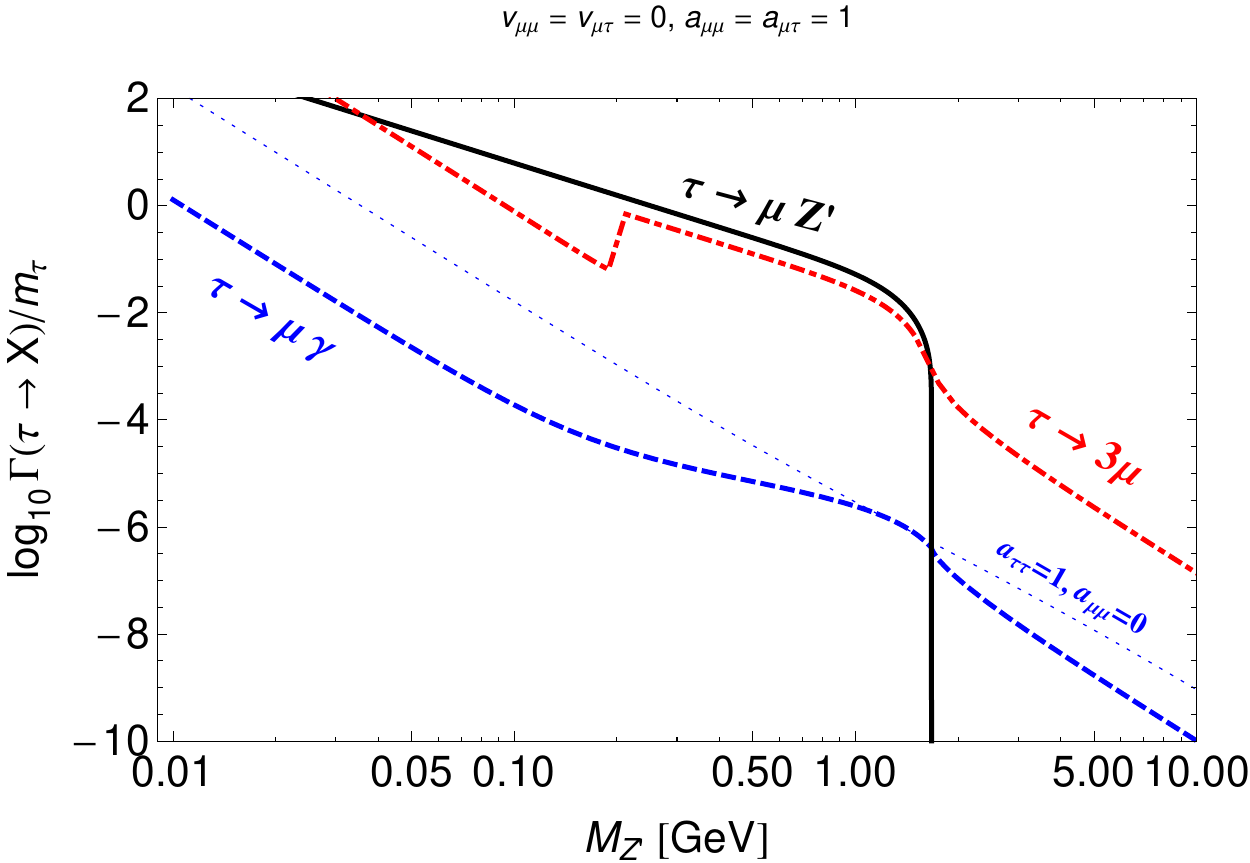}
\caption{$M_{Z'}$ dependence of the decay modes $\tau\to \mu Z'$ (black), $\tau\to 3\mu$ with $\Br (Z'\to\mu\mu)=1/2$ (red, dotdashed), and $\tau\to\mu\gamma$ (blue, dashed) for vector (left) and axial couplings (right). $\tau\to\mu\gamma$ is also shown for $g_{\tau\tau} = 1$, $g_{\mu\mu}=0$ in blue (dotted) for $g=v$ (left) and $g=a$ (right).}
\label{fig:decaymodes}
\end{figure*}

Having identified the preferred region of parameters to explain $(g-2)_\mu$ as $M_{Z'} \simeq \unit[1]{MeV}$--$\unit[1]{GeV}$ with $v_{\mu\mu} = \mathcal{O}(5\times 10^{-4})$, we study the impact of LFV couplings on the model. These arise in particular in $U(1)_{L_\mu-L_\tau}$ models that try to explain $h\to\mu\tau$~\cite{Heeck:2014qea,Crivellin:2015mga,Crivellin:2015lwa}, but are of course of more general interest. For heavy $Z'$, the most constraining LFV decay is typically $\tau\to 3\mu$, with current upper limit $\Br (\tau\to 3\mu) < 1.2\times 10^{-8}$ at $90\%$~C.L.~\cite{Amhis:2014hma}. 
This limit can most likely be improved by an order of magnitude to $10^{-9}$ in the future~\cite{Aushev:2010bq,Alekhin:2015byh}.
For large $M_{Z'}\gg m_\tau\gg m_\mu$, we obtain the well-known expression
\begin{align}
\begin{split}
&\Gamma (\tau \to 3\mu ) \simeq \frac{m_\tau^5}{768 \pi^3 M_{Z'}^4}\\
&\quad\times\left[ 4 v_{\mu\mu} v_{\mu\tau}a_{\mu\mu} a_{\mu\tau} + 3 (v_{\mu\mu}^2 + a_{\mu\mu}^2)(v_{\mu\tau}^2 + a_{\mu\tau}^2)\right] ,
\end{split}
\end{align}
whereas the formulae are much more complicated for small $M_{Z'}$.
For vector interactions, $a_{\mu\mu}=0$, the longitudinal $Z'$ polarization drops out when coupled to the muon current $\overline{\mu}\gamma^\alpha \mu$, so the rate is constant for $M_{Z'} \to 0$ 
\begin{align}
\Gamma (\tau \to 3\mu ) \simeq \frac{m_\tau^3 v_{\mu\mu}^2}{900 \pi^3 m_\mu^2}  (v_{\mu\tau}^2+a_{\mu\tau}^2) + \mathcal{O}\left(\frac{M_{Z'}^2}{m_\tau}\right) ,
\end{align}
but for a non-zero axial current there is an enhancement for small $M_{Z'}$ of the form
\begin{align}
\Gamma (\tau \to 3\mu ) \simeq \frac{a_{\mu\mu}^2 (v_{\mu\tau}^2+a_{\mu\tau}^2) }{128 \pi^3}\frac{m_\tau^3 m_\mu^2}{M_{Z'}^4} \log^2 \left(\frac{m_\mu}{m_\tau}\right)\, ,
\end{align}
for $M_{Z'} \ll m_\mu$.
In the intermediate region $2 m_\mu < M_{Z'} < m_\tau - m_\mu$ the $Z'$ can be produced on-shell, and the rate in the narrow-width approximation takes the form
\begin{align}
\Gamma (\tau \to 3\mu ) \simeq \Gamma (\tau \to \mu Z') \,\Br (Z'\to \mu\mu)\,.
\label{eq:nwa}
\end{align}
Assuming only the decays into $\nu_{\mu,\tau}$ from our effective Lagrangian, we have $\Br (Z'\to \mu\mu) = \mathcal{O}(1)$.
The decay width $\tau\to 3\mu$ is shown in Fig.~\ref{fig:decaymodes} for both vector and axial couplings.
In order for the experimental limits from BaBar and Belle on $\tau \to 3\mu$ to be applicable, we have to require the $Z'\to \mu\mu$ decay to occur well inside the relevant detector, which is the case for the region of interest from Fig.~\ref{fig:muonforce}.
A future discovery of $\tau\to 3\mu$, e.g.~at Belle~II, could reveal the underlying light mediator if enough spectral information can be collected.

The $Z'$ contributes at one loop to $\tau\to\mu\gamma$~\cite{Lavoura:2003xp}, which is constrained to a similar level as $\tau\to 3\mu$~\cite{Amhis:2014hma}. The rate is however suppressed by $\alpha_\mathrm{EM}$ compared to $\tau\to 3\mu$ and thus typically gives much weaker bounds on the underlying parameters. 
Assuming only $v_{\mu\mu}$ and $v_{\mu\tau}$ to be non-zero, the rate for $M_{Z'}\ll m_\mu$ is approximately
\begin{align}
\Gamma (\tau\to\mu\gamma) \simeq \alpha_\mathrm{EM}\frac{v_{\mu\mu}^2 v_{\mu\tau}^2}{128 \pi^3} m_\tau \log^2 \left(\frac{m_\mu}{m_\tau}\right),
\end{align}
see Fig.~\ref{fig:decaymodes}.
While typically much smaller than the $\tau\to 3\mu$ rate, the loop-induced $\tau\to\mu\gamma$ depends also on the coupling $g_{\tau\tau}$, and hence can dominate if $g_{\tau\tau} \gg g_{\mu\mu}$. As a result, $\tau\to\mu \gamma$ is in our setup only useful to obtain limits on $g_{\tau\tau}$ for non-zero $g_{\mu\tau}$.

\begin{figure*}[t]
\includegraphics[width=0.46\textwidth]{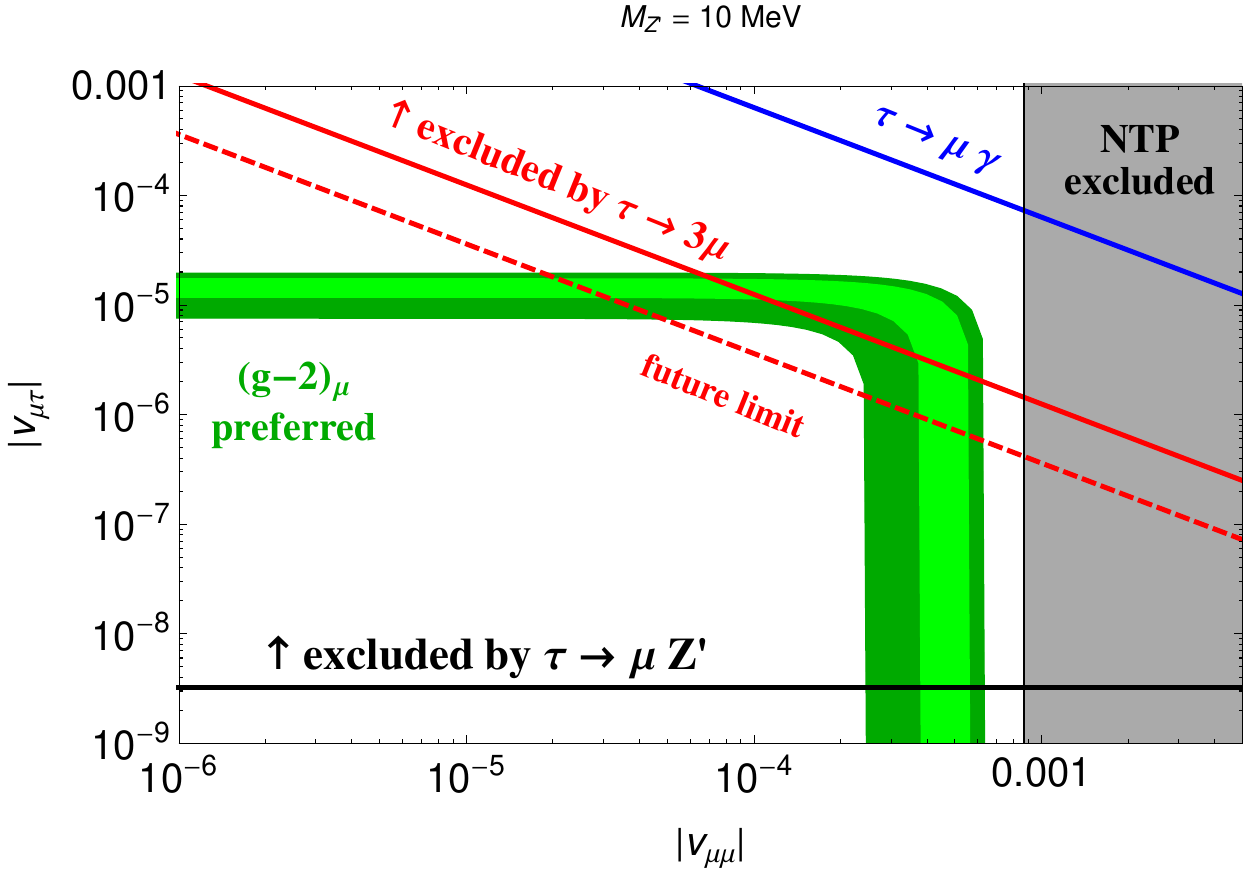}
\includegraphics[width=0.46\textwidth]{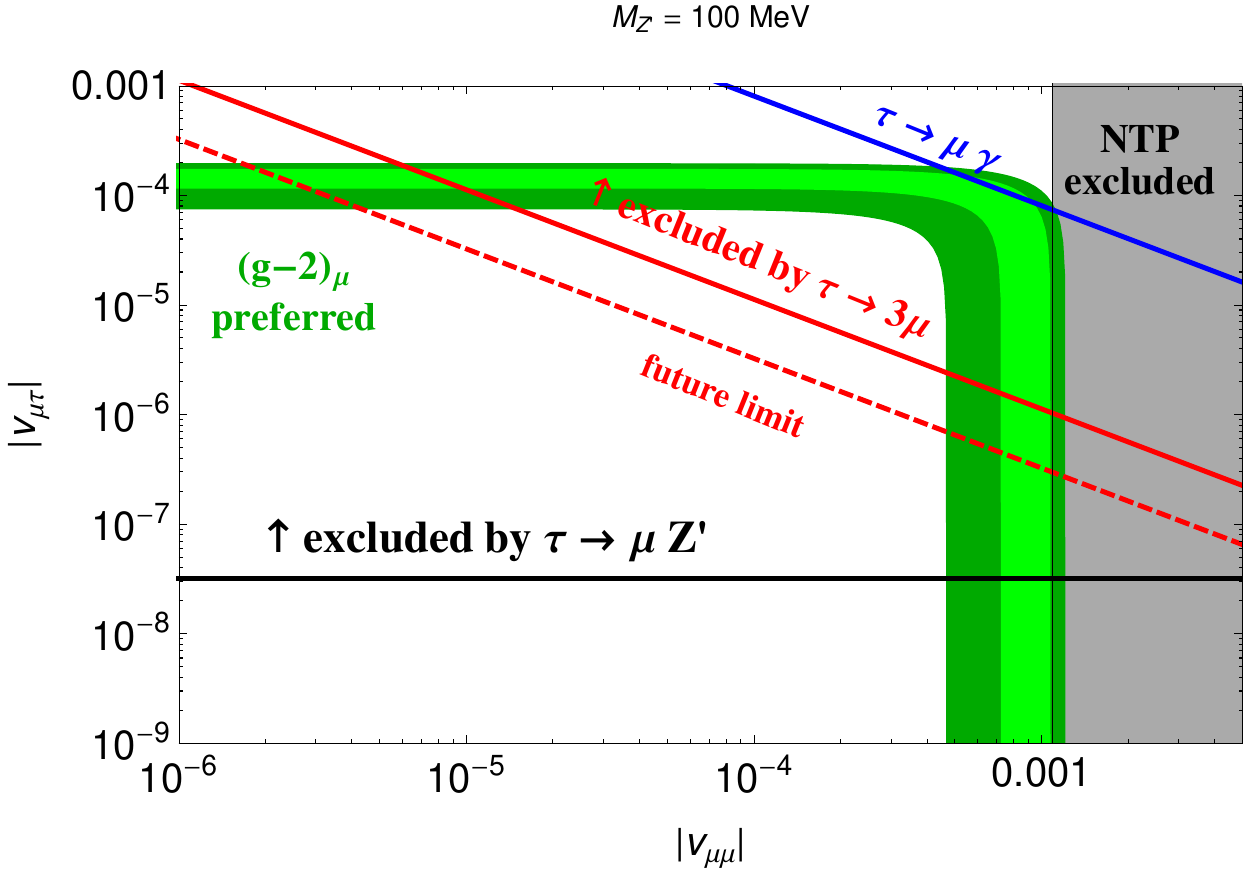}
\caption{Limits on the $Z'$ couplings $v_{\mu\mu,\mu\tau}$ with $M_{Z'} = \unit[10]{MeV}$ (left) and $\unit[100]{MeV}$ (right). The preferred region for $a_\mu$ at ($1\sigma$) $2\sigma$ shown in (light) green, the $95\%$~C.L.~constraint from NTP~\cite{Altmannshofer:2014pba} in gray. Blue and (dashed) red show (future) $90\%$~C.L.~constraints from $\tau\to \mu\gamma$ and $\tau\to 3\mu$, respectively, far surpassed by the $95\%$~C.L.~limit from $\tau\to \mu Z'$~\cite{Albrecht:1995ht} (black).}
\label{fig:limits}
\end{figure*}

As hinted at before, for $M_{Z'} < m_\tau- m_\mu$, which includes our region of interest (Fig.~\ref{fig:muonforce}), the couplings allow for the two-body decay $\tau \to \mu Z'$~\cite{Foot:1994vd,McDonald:2006jr}, followed by $Z'\to \nu\nu$ or $Z'\to \mu\mu$ (see Eq.~\eqref{eq:nwa}). This rarely considered decay turns out to give the best limits on our LFV couplings for $M_{Z'}< 2 m_\mu$. A short calculation gives
\begin{align}
\begin{split}
\Gamma (\tau\to\mu Z')  &= \frac{v_{\mu\tau}^2 }{8\pi}  \frac{m_\tau^2}{M_{Z'}^2} \left[\left(1-\frac{m_\mu}{m_\tau}\right)^2-\frac{M_{Z'}^2}{m_\tau^2}\right] \\
&\quad\times\left[ \left(1+\frac{m_\mu}{m_\tau}\right)^2+\frac{2 M_{Z'}^2}{m_\tau^2}\right] \sqrt{\vec{p}_\mu^2}\,,
\end{split}
\label{eq:tautomuZp}
\end{align}
with muon momentum (squared) defined by
\begin{align}
\vec{p}_\mu^2 = \frac{ \left[ m_\tau^2 - (M_{Z'}+m_\mu)^2\right]\left[ m_\tau^2 - (M_{Z'}-m_\mu)^2\right]}{4 m_\tau^2} \,.
\end{align}
The decay rate via the axial coupling can be obtained from Eq.~\eqref{eq:tautomuZp} by replacing $v_{\mu\tau}\to a_{\mu\tau}$ and $m_\mu \to - m_\mu$.
The rate features a $1/M_{Z'}^2$ enhancement for small $M_{Z'}$ that is known from e.g.~top decays $t\to b W$ and can be understood with help of the Goldstone boson equivalence theorem. In particular it means that the longitudinal $Z'$ polarization dominates in the decay for small $M_{Z'}$.

Experimental constraints on the LFV two-body decay $\tau\to \mu Z'$ with an invisibly decaying $Z'$ have been derived 1995 by ARGUS~\cite{Albrecht:1995ht} (see also older limits in Refs.~\cite{Baltrusaitis:1985fh,Albrecht:1990zj}). Testing seven values of $M_{Z'}$ from $0$ to $\unit[1.6]{GeV}$, limits on $\Br (\tau \to\mu Z')$ at the per-mille level have been obtained, which we will naively interpolate for $Z'$ masses in between. (The limits were actually obtained for pseudo-scalars, but we expect them to be approximately valid for vector bosons as well, especially because the longitudinal part of $Z'$ dominates for small $M_{Z'}$.)
For the sub-GeV masses of interest for us, this leads to the $95\%$~C.L.~limit $\Br (\tau \to\mu Z') \lesssim 5 \times 10^{-3}$~\cite{Albrecht:1995ht}.

Notice that we are looking at $\tau \to \mu + \mathrm{inv}$, which is similar to the SM decay mode $\tau\to\mu \nu\nu$. However, here it is a \emph{two-body} decay, so the resulting muon spectrum is different and allows for an experimental test of this mode (similar to analyses of Michel parameters). This is of course still harder to look for than visible modes such as $\tau\to\mu\gamma$, but Belle and BaBar should be able to improve ARGUS' limit using their $\sim 3000$ times larger set of $\tau$ events. Future $B$ factories such as Belle II can of course improve these limits even further (or discover the decay).
As pointed out long ago in Ref.~\cite{ Goldman:1987hy} (for muon decay), the process $\tau \to \mu \gamma Z'$ can give competitive limits to $\tau\to \mu Z'$ and should also be considered. To our knowledge, there are no limits on $\tau\to\mu \gamma +\mathrm{inv}$ yet.

\section{Discussion}

As discussed above, the relevant region to explain the muon's magnetic-moment anomaly via $v_{\mu\mu}$ requires $Z'$ masses below GeV and thus leads to $\tau\to \mu Z'$ if a LFV $\mu$--$\tau$ coupling $g_{\mu\tau}$ exists. For $M_{Z'} > 2 m_\mu$, the decay $\tau\to 3\mu$ is resonantly enhanced and will generally give the strongest limits on $g_{\mu\tau}$, unless $\Br (Z'\to\mu\mu) \ll 1$.
For $M_{Z'} < 2 m_\mu$---which holds for most of the relevant parameter space---on the other hand, the LFV two-body decay $\tau\to\mu Z'$ followed by $Z'\to \nu\nu$ gives the strongest bounds on $g_{\mu\tau}$, namely $M_{Z'}/|g_{\mu\tau}| > \unit[3.1\times 10^6]{GeV}$ (Fig.~\ref{fig:limits}).

In principle one can also consider dominantly off-diagonal couplings, i.e.~$g_{\mu\mu}\ll g_{\mu\tau}$, as constructed in Ref.~\cite{Foot:1994vd}. A resolution of $(g-2)_\mu$ then requires $M_{Z'}/|v_{\mu\tau}| \simeq \unit[1.4]{TeV}$ with $M_{Z'} >m_\tau-m_\mu$ in order to evade the $\tau\to\mu Z'$ bound. Note that neither the NTP bound nor $\tau\to 3\mu, \mu\gamma$ apply for $g_{\mu\mu,\tau\tau} = 0$, so it is indeed possible to resolve $(g-2)_\mu$ via $v_{\mu\tau}$~\cite{Foot:1994vd}.

Having focused on the $\mu$--$\tau$ sector so far, we will now briefly discuss $Z'$ couplings to electrons, which are much more constrained. Let us replace the taus in Eqs.~\eqref{eq:lagrangian} and~\eqref{eq:nulagrangian} by electrons, so that we still have $v_{\mu\mu}$ to resolve $(g-2)_\mu$ (Fig.~\ref{fig:muonforce}). The electron coupling $g_{ee}$ is required to be far smaller than the muon coupling in order to survive electron experiments~\cite{Lees:2014xha,Batley:2015lha}. The relevant LFV decays are $\mu\to 3 e$, $\mu\to e\gamma$, and, if $M_{Z'} < m_\mu - m_e$, $\mu\to e Z'$. 
$90\%$~C.L.~limits on $\Br (\mu\to e f)$ of order $10^{-5}$~\cite{Bayes:2014lxz} and $10^{-6}$~\cite{Jodidio:1986mz} have been obtained for Majoron-like scalars $f$,
as well as $\Br (\mu\to e \gamma f) < 1.1\times 10^{-9}$ at $90\%$~C.L.~\cite{Goldman:1987hy, Bolton:1988af}.
Since BBN requires at least $M_{Z'} > 2 m_e$, we are however unavoidably in the resonantly enhanced mode of $\mu\to 3 e$ should $\mu\to e Z'$ exist,
\begin{align}
\Br (\mu \to 3 e ) \simeq \Br (\mu \to e Z') \, \Br (Z'\to e e) \,,
\end{align}
which is limited to $10^{-12}$~\cite{Bellgardt:1987du}. For realistic models we expect $\Br (Z'\to e e)$ to be of order one, so the two-body decay $\mu\to e Z'$ can never compete with the well-constrained $\mu\to 3 e$. (A loophole being again purely off-diagonal couplings.)

The third possible combination is a $Z'$ coupling to electrons and taus, leading to the LFV decays $\tau\to 3 e$, $\tau \to e\gamma$, and potentially $\tau\to e Z'$. Without $(g-2)_\mu$ as a guiding principle, we can still conclude that the decay rate $Z'\to ee$ can not be turned off kinematically due to the BBN bound, so it will again be generically impossible for $\tau\to e Z'$ to be observable without violating $\tau \to 3 e$ constraints.
As such, $\tau\to \mu Z'$ is really the standout LFV two-body decay of light new gauge bosons, as it can easily be dominant while respecting existing bounds from e.g.~$\tau\to 3\mu$. Furthermore, the relevant couplings---$v_{\mu\mu}$ and $v_{\mu\tau}$---are motivated by the muon's magnetic moment and the hint for $h\to\mu\tau$.

\section{Conclusion}

If the muon's magnetic-moment anomaly is resolved by a new light gauge boson $Z'$ coupled to muons, its mass is restricted to be between MeV and GeV. Its couplings necessarily violate lepton universality, so it is not far fetched to also assume lepton flavor violation, additionally motivated by the recent hint for $h\to\mu\tau$. We have shown here that the constraints on the LFV $Z'$ couplings do not only come from the usual candidates $\tau\to 3\mu$ or $\tau\to\mu\gamma$, but also from the two-body decay $\tau\to\mu Z'$, courtesy of the small $Z'$ mass, followed by $Z'\to \nu\nu$. 
The current limit on $\tau\to\mu Z'$ comes from the $5\times 10^5$ $\tau$ events studied by ARGUS 20 years ago, and can certainly be improved using Belle's $10^9$ $\tau$ events.

\section*{Acknowledgements}

The author thanks Thomas Hambye for comments on the manuscript and acknowledges support by the F.R.S.-FNRS as a postdoctoral researcher and the use of \texttt{Package-X}~\cite{Patel:2015tea} for calculating loop integrals.

\bibliographystyle{utcaps_mod}
\bibliography{BIB}

\end{document}